\documentclass[a4paper,11pt]{article}
\pdfoutput=1 % if your are submitting a pdflatex (i.e. if you have
\usepackage{jheppub} % for details on the use of the package, please
\usepackage[T1]{fontenc} % if needed
\usepackage{tikz-cd}
\usepackage{caption}
\title{\boldmath  Root-$T\bar{T}$ Deformations  on Causal Self-Dual  Electrodynamic  Theories}

\author[a,b]{H. Babaei-Aghbolagh,}
\author[c]{Komeil Babaei Velni,}
\author[a,b,d]{Song He,}
\author[c]{Zahra Pezhman}
\affiliation[a]{Institute of Fundamental Physics and Quantum Technology,  Ningbo University, Ningbo, Zhejiang 315211, China}
\affiliation[b]{School of Physical Science and Technology, 
	Ningbo University, Ningbo, 315211, China}
\affiliation[c]{Department of Physics, University of Guilan, P.O. Box 41335-1914, Rasht, Iran}
\affiliation[d]{Max Planck Institute for Gravitational Physics (Albert Einstein Institute),
	Am M\"uhlenberg 1, 14476 Golm, Germany}
\emailAdd{hosseinbabaei@nbu.edu.cn}
\emailAdd{babaeivelni@guilan.ac.ir}
\emailAdd{ hesong@nbu.edu.cn}
\emailAdd{ zpezhmanh@phd.guilan.ac.ir}

\abstract{The self-dual condition, which ensures invariance under electromagnetic duality, manifests as a partial differential equation in nonlinear electromagnetism theories. The general solution to this equation is expressed in terms of an auxiliary field, $\tau$, and Courant-Hilbert functions, $\ell(\tau)$, which depend on $\tau$. Recent studies have shown that duality-invariant nonlinear electromagnetic theories fulfill the principle of causality under the conditions $\frac{\partial \ell}{\partial \tau} \ge 1$ and $\frac{\partial^2 \ell}{\partial \tau^2} \ge 0$.
	
	In this paper, we investigate theories with two coupling constants that also comply with the principle of causality. We demonstrate that these theories possess a new universal representation of the root-$T\bar{T}$ operator. Additionally, we derive marginal and irrelevant flow equations for the logarithmic causal self-dual electrodynamics.}

\begin{document}
\maketitle
\flushbottom
	
%%%%%%%%%%%%%%%%%%%%%%%%%%%%%%%%%%%%%%%%%%%%%%%%%%%
\section{Introduction}\label{0} 
%%%%%%%%%%%%%%%%%%%%%%%%%%%%%%%%%%%%%%%%%%%%%%%%%%%%
The Protection of $SO(2)$ duality symmetry is an essential feature of extended models of nonlinear electrodynamics (NED) \cite{Born:1933lls, Born:1934gh, Plebanski, Boillat, Sorokin:2021tge}. This symmetry, inherent in Maxwell's equations, ensures that the electric and magnetic fields can transform into each other through rotations in the field space \cite{Bialynicki-Birula:1981, Bialynicki-Birula:1992rcm, Gaillard:1981rj, Gaillard:1997rt, Gibbons:1995ap}. When Maxwell's theory is deformed while preserving this symmetry, it gives rise to a rich structure of NEDs, which may have significant implications for the effective action of D-branes \cite{Fradkin:1985qd, Garousi:2017fbe, Babaei-Aghbolagh:2013hia}.

The Gaillard-Zumino approach is a foundational element in the theory of self-dual nonlinear electrodynamics models \cite{Gaillard:1981rj, Gaillard:1997rt}. This method provides a systematic way to construct invariant interaction terms that remain unchanged under duality transformations. The original work by Gaillard and Zumino established the groundwork, which was further elaborated in \cite{Gibbons:1995ap} by Gibbons and Rasheed. The non-covariant Hamiltonian approach was initially introduced by Henneaux and Teitelboim \cite{HT} and further developed by Deser and his colleagues \cite{Deser:1997mz}. This alternative perspective employs Hamiltonian formalism to identify constraints and quantities invariant under duality transformations while maintaining non-manifest Lorentz invariance. This framework offers a distinctive standpoint on self-duality \cite{Deser:1997se}. Finally, the PST method, developed by Pasti, Sorokin, and Tonin, is another significant technique in this field. This approach employs auxiliary fields and a covariant formulation to achieve duality invariance. The PST method has proven to be instrumental in constructing actions for various field theories that exhibit self-duality \cite{Mkrtchyan:2205uvc, Mkrtchyan:2019opf}. In Gaillard-Zumino approach, the Lagrangian of  duality invariant  nonlinear electrodynamics models satisfies the self-duality condition \cite{Bialynicki-Birula:1981, Bialynicki-Birula:1992rcm}(see also \cite{Gibbons:1995ap, Gaillard:1997rt, Kuzenko:2000uh, Aschieri:2008ns, Bossard:2011ij, Carrasco:2011jv, Chemissany:2011yv, Aschieri:2013nda}):
\begin{equation}
	G\tilde{G}+F\tilde{F}=0\,,
\end{equation}
where $\tilde{G}$ is an antisymmetric tensor field defined as $\tilde{G}^{\mu \nu} = 2 \frac{\partial {\cal L}(F)}{\partial F_{\mu \nu}}$. The self-duality condition, in its differential form, can be expressed through the following partial differential equation (PDE):
\begin{eqnarray}
	\label{lagrangeNGZ}
	(\partial_S \mathcal{L})^2\, -2 \frac{S}{P} \, (\partial_S \mathcal{L}) (\partial_P \mathcal{L})-(\partial_P \mathcal{L})^2\, =1\,,
\end{eqnarray}
where $S=-\frac{1}{4}F_{\mu\nu}F^{\mu\nu}$ and $\,\,P=-\frac{1}{4}F_{\mu\nu}\tilde F^{\mu\nu}$ are two Lorentz invariant quantities. The condition~\eqref{lagrangeNGZ} guarantees that the equations of motion derived from the Lagrangian remain invariant under the exchange of electric and magnetic fields, embodying the fundamental symmetries central to non-linear electrodynamics theories \cite{Kallosh:2011dp}. To simplify the self-duality equation~\eqref{lagrangeNGZ}, the Lagrangian 
$\mathcal{L}$ can be parametrized in terms of two alternative non-negative independent variables, $V$ and $U$,
\begin{eqnarray}
	\label{UN}
	U=\tfrac12(\sqrt{S^2+P^2}-S)\, ,\quad V=\tfrac12(\sqrt{S^2+P^2}+S)\, .
\end{eqnarray}
The resulting equation is \cite{Gaillard:1997rt}
\begin{eqnarray}
	\label{dUdN}
	\partial_U \mathcal{L}\,\, \partial_V \mathcal{L}=-1\,.
\end{eqnarray}
The self-dual condition, corresponding to equation~\eqref{dUdN}, leads to an infinite number of solutions for non-linear electrodynamic Lagrangians. A general method for solving the differential equation~\eqref{dUdN} is discussed in \cite{cour}. The solution that corresponds to the initial condition \(\mathcal{L}(0, V) = \ell(V)\) is given by:
\begin{eqnarray}
	\label{SPDE}
	\mathcal{L}=\ell (\tau)-\frac{2U}{\dot\ell(\tau)}\ , 
	\qquad  \tau=V+\frac{U}{\dot\ell^2(\tau)}\, \end{eqnarray}
In the following, we shall refer to equation~\eqref{SPDE} as the Courant-Hilbert (CH) solution and denote $\ell(\tau)$ as the ``CH-function". By definition, $\tau$ is always non-negative, becoming zero only when $U$ and $V$ are zero. When $\dot{\ell}(\tau) > 0$, the Lagrangian~\eqref{SPDE} satisfies the differential equation~\eqref{dUdN}. The specific choice $\ell(\tau) = \tau$ corresponds to the free-field Maxwell theory \cite{Russo2024causal}. The energy-momentum tensor (EMT) associated with the Lagrangian~\eqref{SPDE} is \cite{Russo:2024xnh, Russo2024dual}
\begin{equation}\label{TMNl}
	T_{\mu\nu} = \left(\frac{\tau\dot\ell}{U+V}\right) T^{\rm Max}_{\mu\nu} +
	(\ell - \tau \dot\ell) {\rm g}_{\mu\nu} \, ,  
\end{equation}
where $T^{Max}_{\mu\nu}=F_{\mu\rho}{F_{\nu}}^{\rho}-\frac{1}{4} \,g_{\mu\nu} F_{\alpha\beta}F^{\alpha\beta}$ represents the stress-energy tensor (EMT) of Maxwell's theory. The trace of Maxwell's stress-energy tensor vanishes. Consequently, the trace of the EMT in any self-dual electrodynamic theory can generally be written as follows \cite{Russo:2024xnh}:
\begin{equation}\label{TnNl}
	{T_{\mu}}^\mu = 4(\ell - \tau \dot\ell)  \, .  
\end{equation}
Therefore, the traceless condition for a self-dual electrodynamic theory is satisfied when the equation $\ell= \tau \dot\ell$ is fulfilled. Including interaction terms in a two-dimensional free scalar theory, utilizing an irrelevant operator for scalar theories is discussed in \cite{Cavaglia:2016oda, Smirnov:2016lqw}. This operator is a specific function of the momentum-energy tensor, expressed as $O_{\lambda}=\frac{1}{8} \Big(T_{\mu\nu}T^{\mu\nu}  - {T_{\mu}}^{\mu} {T_{\nu}}^{\nu}\Big)$. The two-dimensional marginal operator study for these theories can also be found in \cite{Conti:2022egv, Babaei-Aghbolagh:2022kjj, Ferko:2206jsw}. A systematic review of the existing literature on $T\bar{T}$ deformations is provided in \cite{Jiang19, Song2025}. A recent development in 4D field theory is the addition of both irrelevant operator (called $T \bar{T}$-like deformation \cite{Conti:2018jho}) and marginal  operator (called root $T \bar{T}$ deformation \cite{Babaei-Aghbolagh:2022MoxMax})  to the free Lagrangian (Maxwell theory), which results in a deformed Lagrangians of the form:  
\begin{equation}
	\label{L1}
	{\cal L}_{\lambda}={\cal L}_{Max} +\int O_\lambda d\lambda \,\,,\qquad\,\,{\cal L}_{\gamma}={\cal L}_{Max} +\int {\cal R}_\gamma d\gamma ,
\end{equation}
where  $\lambda$ and $ \gamma $ are two dimensionful and dimensionless coupling parameters, the deformation operator  $O_{\lambda}$ and ${\cal R}_\gamma$ is a special function of the (EMT) of the seed theory. Thus, the two flow equations for the coupling of $\lambda$ and $\gamma$ can be expressed as:
\begin{eqnarray}
	\frac{\partial \mathcal{L}}{\partial \lambda} =O_\lambda\,\,, \qquad \,\,\frac{\partial \mathcal{L}}{\partial \gamma} = {\cal R}_\gamma\,,
\end{eqnarray} 
within the context of the given deformation theory. Generally, the two operators $O_\lambda$ and ${\cal R}_\gamma$ are functions of the (EMT). These functions comprise two independent structures, ${T_{\mu}}^{\mu} {T_{\nu}}^{\nu}$ and $T_{\mu\nu}T^{\mu\nu}$.
By choosing the irrelevant deformation operator \cite{Conti:2018jho}:
\begin{equation}
	O_{\lambda}=\frac{1}{8} \Big(T_{\mu\nu}T^{\mu\nu}  -\frac{1}{2} {T_{\mu}}^{\mu} {T_{\nu}}^{\nu}\Big),
\end{equation}  and using the perturbation approach the extra terms in Maxwell’s theory correspond precisely to the expansion of the Born-Infeld Lagrangian as :
\begin{eqnarray}
	\label{LEBI}
	{\cal L}_{BI}&=&{\cal L}_{Max} +\int O_\lambda d\lambda \nonumber\\
	&=& S + \tfrac{1}{2} \lambda (S^2 +P^2)+  \tfrac{1}{2}S\lambda^2 (S^2 +P^2)+\dots\,\, \nonumber\\
	&=&\frac{1}{\lambda} \bigg( 1 -  \sqrt{1- 2 \lambda S-\lambda^2 P^2 } \bigg)=  \frac{1}{\lambda} - \sqrt{\left(\frac{1}{\lambda}+ 2U\right)\left(\frac{1}{\lambda} -2 V\right)} \,.
\end{eqnarray}
The Born-Infeld Lagrangian is a theory that meets the differential self-duality condition in the PDE~\eqref{dUdN} and features an irrelevant flow equation represented by $\frac{\partial \mathcal{L}_{BI}}{\partial \lambda} =\frac{1}{8} \Big(T_{\mu\nu}T^{\mu\nu}  -\frac{1}{2} {T_{\mu}}^{\mu} {T_{\nu}}^{\nu}\Big)$. The CH-function for  Born-Infeld Lagrangian is $\ell (\tau) = \frac{1}{\lambda}- \sqrt{\frac{1}{\lambda}\left(\frac{1}{\lambda}-2\tau\right)}$ \cite{Russo2024causal}.

The ModMax theory has a marginal deformation parameter ($\gamma$) that preserves the conformal symmetry and the gauge invariance of the Maxwell theory\cite{Bandos:2020jsw}. In a recent work \cite{Babaei-Aghbolagh:2022MoxMax}, we introduced the ModMax theory as a root-type $T \bar{T}$ deformation of Maxwell's theory.
The root-type $T \bar{T}$ operator in four dimensions is:
\begin{eqnarray}\label{Ogama}
	{\cal R}_{\gamma}= \frac{1}{2}
	\sqrt{T_{\mu\nu}T^{\mu\nu}- \frac{1}{4} {T_{\mu}}^{\mu} {T_{\nu}}^{\nu}}\,.
\end{eqnarray}
In ModMax theory, the CH-function is represented in \cite{Russo2024causal} by $\ell (\tau) =e^\gamma \tau$, and the Lagrangian density depends on two variables, $V$ and $U$, as follows:
\begin{equation}
	\label{LMM}
	{\cal L}_{MM}=e^\gamma V- e^{-\gamma}U.
\end{equation}
The ModMax theory presents a distinctive Lagrangian in non-linear electrodynamics, maintaining conformal and electromagnetic-duality invariance. Consequently, ModMax theory is traceless, and as indicated by Eq.~\eqref{TnNl}, we have $\ell= \tau \dot\ell$ for this theory.

Ref. \cite{Babaei-Aghbolagh:2022MoxMax} demonstrates that a root-type $T \bar{T}$ like deformation facilitates the transformation of the BI theory into the Generalized Born-Infeld (GBI) theory (a BI-type deformation of ModMax \cite{Bandos:2020hgy}). The  Lagrangian density and CH-function of this theory are denoted by:
\begin{equation}
	\label{MMB}
	\quad  \mathcal{L}_{GBI} =  \frac{1}{\lambda} - \sqrt{\left(\frac{1}{\lambda}+ 2e^{-\gamma}U\right)\left(\frac{1}{\lambda} -2e^\gamma V\right)} \, \,\,,\qquad\,\,\ell (\tau) = \frac{1}{\lambda}- \sqrt{\frac{1}{\lambda}\left(\frac{1}{\lambda}-2e^{\gamma}\tau\right)}\,\,. 
\end{equation}
Refs. \cite{Babaei-Aghbolagh:2022MoxMax, Ferko:2022iru, Aghbolagh2210,ferko2024interacting} indicates that, with respect to the $\lambda$ and $\gamma$ couplings, there are two flow equations considered irrelevant and marginal. These equations are expressed as follows:
$\partial {\cal L}_{GBI}/{\partial \gamma} = \frac{1}{2}
\sqrt{T_{\mu\nu}T^{\mu\nu}- \frac{1}{4} {T_{\mu}}^{\mu} {T_{\nu}}^{\nu}}$ and  $\partial {\cal L}_{GBI}/{\partial \lambda} = \frac{1}{8} \Big(T_{\mu\nu}T^{\mu\nu}  -\frac{1}{2} {T_{\mu}}^{\mu} {T_{\nu}}^{\nu}\Big)$.

For self-dual (NLED) theories in the weak-field limit, it was demonstrated in \cite{Russo2024causal} that the causality conditions simplify to the following inequalities involving the derivatives of the CH-function $\ell(\tau)$:
\begin{equation}
	\label{CausalCondition}
	\dot\ell \ge 1 \, , \qquad \ddot \ell\ge0 \, .  
\end{equation}
This finding is noteworthy for its simplicity and because there was no initial reason to expect that the causality conditions on $\ell$ would be independent of the variables $(U, V)$. Additionally, the condition $\ddot\ell \ge 0$ indicates that $\ell(\tau)$ is a convex function\cite{Russo2024causal}.

In this paper, we explore the irrelevant and marginal flow equations of electrodynamic theories that adhere to the principles of causality and duality. We focus on theories that satisfy the PDE condition in~\eqref{dUdN} and also meet condition~\eqref{CausalCondition}. The Born-Infeld, ModMax, and General Born-Infeld theories are examples of these flow equations that were previously examined.
Additionally, recent studies by Russo and Townsend have investigated two other significant theories, which involve the $\gamma$ and $\lambda$ couplings \cite{Russo2024causal, Russo2024dual}. One is a logarithmic theory, while the other is a no maximum-$\tau$ theory \cite{Russo2024dual}. We will analyze the flow equations of  Logarithmic self-dual electrodynamics, demonstrating that Logarithmic self-dual theory exhibits marginal flow equations with respect to the $\gamma$ coupling constant and irrelevant flow equations with respect to the $\lambda$ coupling constant. %Furthermore, we recognize a symmetry, referred to as $\alpha$-symmetry, which applies to all self-dual electrodynamic theories discussed in this paper.

%%%%%%%%%%%%%%%%%%%%%%%%%%%%%%%%%%%%%%%%%%%%%
\section{New representation of the   root $T\bar{T}$-operator }
\label{3.0}
%%%%%%%%%%%%%%%%%%%%%%%%%%%%%%%%%%%%%%%%%%%%%%%%%%%%%%%%%%%%%%%%%%%%%%%%%%%%%%%%%%%%
The root $T\bar{T}$ operator $R_\gamma$ in Eq.~\eqref{Ogama} is a special function of two independent structures, ${T_{\mu}}^{\mu} {T_{\nu}}^{\nu}$ and $T_{\mu\nu}T^{\mu\nu}$. These structures can be expressed using Eq.~\eqref{TMNl} as follows \cite{Russo:2024xnh}:
\begin{equation}\label{Tl}   
	{T_{\mu}}^{\mu} {T_{\nu}}^{\nu}= 16(\ell - \tau\dot\ell)^2\, ,   
	\qquad
	T_{\mu\nu}T^{\mu\nu}  =  4\left[(\tau\dot\ell)^2 + (\ell -\tau\dot\ell)^2\right] \, .  
\end{equation}
By substituting the values of the introduced structures ${T_{\mu}}^{\mu} {T_{\nu}}^{\nu}$ and $T_{\mu\nu}T^{\mu\nu}$ from Eq.~\eqref{Tl} into operator $R_\gamma$ in~\eqref{Ogama}, we can derive a new representation for operator $R_\gamma$. This  representation is as follows:
\begin{eqnarray}\label{Rtua}
	\mathcal{R}_\gamma &=& \tau\dot\ell\,.
\end{eqnarray}
The operator~\eqref{Rtua} is a unique relation for all duality Lagrangians obtained from the CH-function in Eq.~\eqref{SPDE}.
This section explores electrodynamic theories characterized by a marginal flow equation for $\gamma$ coupling. These theories simplify to ModMax theory in the weak field limit. According to the value of the marginal operator $R_\gamma$ in Eq.~\eqref{Rtua}, we claim that all theories of causal electromagnetism must satisfy the following flow equation:
\begin{eqnarray}\label{Rootf}
	\frac{\partial \mathcal{L}}{\partial \gamma} &=& \tau\dot\ell\,.
\end{eqnarray}
We have already seen this flow equation for the general ModMax and Born-Infeld theories. Here, as a clear example, we study the flow equation~\eqref{Rootf} for the general Born-Infeld model.
%%%%%%%%%%%%
\subsubsection*{ General Born-Infeld Theory: An example}
%%%%%%%%%%%%%%
It can be explicitly verified that the general Born-Infeld Lagrangian and the CH-function in~\eqref{MMB} satisfy the root flow equation~\eqref{Rootf}.
By taking the derivative of the CH-function~\eqref{MMB} with respect to $\tau$, we obtain $\dot\ell$ for the general Born-Infeld theory as given in:
\begin{eqnarray}\label{RootGBIi}
	\dot\ell&=& \frac{e^{\gamma}}{\sqrt{1 - 2 e^{\gamma} \lambda \tau}}\,.
\end{eqnarray}
By substituting $\dot\ell$ from the equation above into the second equation~\eqref{SPDE}, we obtain the following:
\begin{eqnarray}\label{RootGBIi}
	\tau&=&V + e^{-2 \gamma} (U - 2\,  \lambda\, e^{\gamma}\, U \,\,\tau)\,.
\end{eqnarray}
The value of $\tau$ for the GBI theory is determined by solving  above equation as follows: $\tau=\frac{U + e^{2 \gamma} V}{e^{\gamma} (2 \lambda U + e^{\gamma}) }$. Therefore, the right side of flow equation~\eqref{Rootf} for the GBI theory is obtained:
\begin{eqnarray}\label{RootGBi}
	\tau\dot\ell&=& \frac{U + e^{2 \gamma} V}{(2 \lambda U + e^{\gamma}) \sqrt{\frac{ e^{\gamma}-2 e^{2 \gamma} \lambda V }{2 \lambda U + e^{\gamma} }} }\,.
\end{eqnarray}
The left side of the root flow equation~\eqref{Rootf} for the GBI theory can be directly calculated from the Lagrangian's derivative with respect to $\gamma$. Consequently, it can be demonstrated that the root flow equation is valid for the GBI theory. Therefore, we have:
\begin{eqnarray}\label{DgaGBI}
	\frac{\partial \mathcal{L}_{GBI}}{\partial \gamma} &=&	\tau\dot\ell\,.
\end{eqnarray}
%%%%%%%%%%%%%%%%%%%%%%%%%%%%%%%%%%%%%%%%%%%%%%%%%%%
\subsection{Root-$T\bar{T}$ deformations for all causal theories}\label{222.11} 
%%%%%%%%%%%%%%%%%%%%%%%%%%%%%%%%%%%%%%%%%%%%%%%%%%%%
For all causal theories, the CH function can incorporate higher orders of $\tau$ with a dimensionful coupling constant of  $\lambda$. One notable property of the function $\ell(\tau)$ is its power series expansion around $\tau = 0$, ensuring that:
\begin{equation}\label{expan}
	\ell (\tau) = e^\gamma \tau + \lambda \, \mathcal{O}(\tau^2) \, . 
\end{equation}
All theories with double coupling constants  $\gamma$ and $\lambda$, derived from CH functions with the property described in~\eqref{expan}, reduce to the ModMax theory in the limit $\lambda= 0$. 
Our proposal suggests that any theory with expansion
~\eqref{expan} will satisfy~\eqref{Rootf}. If our proposal is correct, we should be able to demonstrate that by including higher orders of $\tau$ in an expansion with arbitrary coefficients,  root flow equation~\eqref{Rootf} still holds.\footnote{We would like to thank Jorge G. Russo for discussions on this point.}
More precisely, let us consider the following expansion:
\begin{eqnarray}\label{ExplT}
	\ell (\tau) =e^\gamma \,  \tau + m_1 \, \lambda \,  e^{2\gamma} \, \tau^2 +m_2\, \lambda^2 \,  e^{3\gamma} \, \tau^3+m_3\, \lambda^3 \,  e^{4\gamma} \, \tau^4  +... 
\end{eqnarray}
where $m_1$
and $m_2$ are two constants.
Using the above expansion and the second equation~\eqref{SPDE}, we can obtain the auxiliary field, $\tau$, up to the $\lambda^3$ order as follows:
\begin{eqnarray}\label{Gq1q22}
	\tau&=&e^{-2 \gamma} U + V - 4\,  m_1 \, \lambda e^{-3 \gamma} U (U + e^{2 \gamma} V)+ \lambda^2 e^{-4 \gamma} \bigl( 4 m_1^2 U (U + e^{2 \gamma} V) (7 U + 3 e^{2 \gamma} V)\nonumber\\
	&&-6 m_2 U (U + e^{2 \gamma} V)^2 \bigr)+4 \lambda^3 e^{-5 \gamma} U \Bigl(-2 m_3 (U + e^{2 \gamma} V)^3 -  m_1 (U + e^{2 \gamma} V) \nonumber\\
	&&\times\bigl(-9 m_2 (U + e^{2 \gamma} V) (3 U + e^{2 \gamma} V)+ 4 m_1^2 (5 U + e^{2 \gamma} V) (3 U + 2 e^{2 \gamma} V)\bigr)\Bigr) .  
\end{eqnarray}
Using the first equation~\eqref{SPDE} and the $\tau$ in equation~\eqref{Gq1q22}, we can obtain the Lagrangian in terms of the two variables $U$ and $V$. By substituting $U$ and $V$ from~\eqref{UN}, we will have the Lagrangian of the order of $\lambda^3$ in terms of the two variables $S$ and $P$ as follows:
\begin{eqnarray}\label{Gq1q2n11}
	&&\mathcal{L} = S \cosh(\gamma) + \sqrt{P^2 + S^2} \sinh(\gamma) + m_1 \lambda \bigl(\sqrt{P^2 + S^2} \cosh(\gamma) + S \sinh(\gamma)\bigr)^2 \\
	&&+\lambda^2 \Bigl(-2 e^{-\gamma} m_1^2 \bigl(- S + \sqrt{P^2 + S^2}\bigr) \bigl(\sqrt{P^2 + S^2} \cosh(\gamma) + S \sinh(\gamma)\bigr)^2 \nonumber\\
	&&+ m_2 \bigl(\sqrt{P^2 + S^2} \cosh(\gamma) + S \sinh(\gamma)\bigr)^3\Bigr) +\lambda^3 \biggl(\tfrac{1}{2} e^{-4 \gamma} m_1^3 \Bigl((1 + e^{2 \gamma})^2 (3 + e^{2 \gamma}) P^4\nonumber\\
	&& + 24 S^3 \bigl(S - \sqrt{P^2 + S^2}\bigr) + 2 P^2 S \bigl((12 + 7 e^{2 \gamma} + e^{6 \gamma}) S + (-6 - 7 e^{2 \gamma} + e^{6 \gamma})\sqrt{P^2 + S^2}\bigr)\Bigr) \nonumber\\
	&&-  \frac{e^{-4 \gamma}}{4 }3 m_1 m_2 \Bigl((1 + e^{2 \gamma})^3 P^4 + 8 S^3 \bigl(S - \sqrt{P^2 + S^2}\bigr)\nonumber\\
	&& + 2 (1 + e^{2 \gamma}) P^2 S \bigl(4 S + (-2 + e^{2 \gamma}) (1 + e^{2 \gamma})\sqrt{P^2 + S^2} + 2 e^{3 \gamma} S \sinh(\gamma)\bigr)\Bigr) + \nonumber\\
	&&\frac{1}{8} m_3 \Bigl(3 P^4 + 4 P^2 (P^2 + 2 S^2) \cosh(2 \gamma) + (P^4 + 8 P^2 S^2 + 8 S^4) \cosh(4 \gamma) \nonumber\\
	&&+ 8 S\sqrt{P^2 + S^2} \bigl(P^2 + (P^2 + 2 S^2) \cosh(2 \gamma)\bigr) \sinh(2 \gamma)\Bigr)\biggr)\nonumber\,.  
\end{eqnarray}
%%%%%%%%%%%%%%%%%%%%%%%%%%%%%%%%%%%%%%%%%%%%%%%%%%%%%%%%%%%%%%%%%%%%%%%%%%%%%%%%%%%%%%%%%%%%%%%%%%%%%%%%%%%%%%%%%%%%%%%%%%%%%%%%%%%%%%%%%%%%%%%%%%%%%%%%%%%%%%%%%%%
If we show that for Lagrangian~\eqref{Gq1q2n11}, the flow equation~\eqref{Rootf} holds independently of the $m_i$ coefficients, then we have proven that the flow equation ~\eqref{Rootf}  holds for all causal theories with different $m_i$. For this purpose, we can obtain the right-hand side of the flow equation~\eqref{Rootf} by expanding the CH-function by multiplying it by the auxiliary field in equation~\eqref{Gq1q22}. We can also obtain the left-hand side of the flow equation~\eqref{Rootf} by differentiating the Lagrangian~\eqref{Gq1q2n11} with respect to $\gamma$. In this case, the flow equation~\eqref{Rootf} will be as follows:
\begin{eqnarray}
	\label{DL12k}
	\frac{\partial \mathcal{L}}{\partial \gamma}&=&\tau\dot\ell\\
    &=&e^{-\gamma} U + e^{\gamma} V - 2  \lambda e^{-2 \gamma} m_1 ( U^2 - e^{4 \gamma} V^2) + \lambda^2 e^{-3 \gamma} (U + e^{2 \gamma} V) \bigl(4 m_1^2 U (3 U -  e^{2 \gamma} V)\nonumber \\
   &- &3 m_2 (U^2 -  e^{4 \gamma} V^2)\bigr) +\lambda^3  e^{-4 \gamma} \bigl(-4 (24 m_1^3 - 12 m_1 m_2 + m_3) U^4 \nonumber\\
   &- & 8 e^{2 \gamma} (14 m_1^3 - 9 m_1 m_2 + m_3) U^3 V + 8 e^{6 \gamma} (2 m_1^3 - 3 m_1 m_2 + m_3) U V^3 + 4 e^{8 \gamma} m_3 V^4\bigr) \,\nonumber
\end{eqnarray}
In the root-flow equation above, the coefficients of $m_i$ can take arbitrary values, yet flow equation $\frac{\partial \mathcal{L}}{\partial \gamma}=\tau\dot\ell$ holds for all such values. Moreover, it holds explicitly for all causal theories up to order $\lambda^3$.
On the other hand, we can show explicitly that the Lagrangian~\eqref{Gq1q2n11} satisfies the following root flow equation up to order $\lambda^3$:
\begin{equation}
	\label{DLBITgp}
	\frac{\partial \mathcal{L}}{\partial \gamma}=\frac{1}{2}\, \sqrt{T_{\mu\nu}T^{\mu\nu}-\frac{1}{4} {T_{\mu}}^{\mu} {T_{\nu}}^{\nu}}=\tau\dot\ell\,
\end{equation}

All causal theories up to any order of $\lambda$ can be obtained from the following CH-function expansion:
\begin{equation}\label{Lexpan}
	\ell (\tau) =e^\gamma \,  \tau  +\sum_{i=1}^\infty m_i \, \lambda^i\, e^{(i+1)\gamma} \, \tau^{i+1} 
\end{equation}
Given a suitable set of $m_i $ coefficients, we can generate a CH-function expansion of the theory up to any order of $\lambda$.
The Lagrangian corresponding to each order  $\lambda$ expansion of~\eqref{Lexpan} applies to the root flow equation~\eqref{Rootf} up to that order of $\lambda$ expansion.
%%%%%%%%%%%%%%%%%%%%%%%%%%%%%%%%%%%%%%%%%%%%%%%%%%%
\subsection{Obtaining the CH-function from the root flow equation}\label{22.222} 
%%%%%%%%%%%%%%%%%%%%%%%%%%%%%%%%%%%%%%%%%%%%%%%%%%%%
The root flow equation~\eqref{Rootf} can be written as a differential equation with derivatives with respect to the two parameters $\gamma$ and $\tau$ and solved with respect to these two variables.
For this purpose, by substituting the Lagrangian in Eq.~\eqref{SPDE} into the flow equation~\eqref{Rootf}, we can form a differential equation as follows:
\begin{equation}
	\partial_\gamma \ell + \frac{U}{\partial_\tau \ell^2}\partial_\tau \partial_\gamma \ell =\partial_\gamma \ell +(\tau-V)\partial_\tau \partial_\gamma \ell = \tau\partial_\tau \ell\,.
\end{equation}
The above PDE can be solved by separation of variables. Assuming
\begin{equation}
	\ell(\tau, \gamma)=T(\tau) G(\gamma)\,,
\end{equation}
we obtain
\begin{equation}
	T(\tau) G^{\prime}(\gamma)+(\tau-V) T^{\prime}(\tau) G^{\prime}(\gamma)=\tau T^{\prime}(\tau) G(\gamma)\,,
\end{equation}
which yields, after some algebraic manipulations,
\begin{equation}
	\frac{G^{\prime}(\gamma)}{G(\gamma)}=\frac{\tau T^{\prime}(\tau)}{T(\tau)+(\tau-V) T^{\prime}(\tau)} \,.
\end{equation}
Since the left side depends only on $\gamma$ and the right side only on $\tau$, both sides must equal a constant, say $\xi$. That is, we have the two ordinary differential equations (ODEs):
\begin{equation}
	\frac{G^{\prime}(\gamma)}{G(\gamma)}=\xi\,, \quad \frac{\tau T^{\prime}(\tau)}{T(\tau)+(\tau-V) T^{\prime}(\tau)}=\xi\,.    
\end{equation}
The ODE for $G(\gamma)$ is straightforward, and it integrates to
\begin{equation}
	G(\gamma) \propto \exp(\xi\gamma)\,.
\end{equation}
On the other hand, we have
\begin{equation}
	T(\tau) \propto \begin{cases}
		\left((1-\xi) \tau+\xi V\right)^{\frac{\xi}{1-\xi}}\,\, \text{ if } \xi \in \mathbb{R}\smallsetminus\{1\}\,,\\
		\exp(\tau/V)\,\, \text{ if } \xi = 1\,.
	\end{cases}
\end{equation}
A general solution can be obtained by superposition:
\begin{equation}
	\ell = \int_{\mathbb{R}\smallsetminus\{1\}}  d \xi \, c(\xi) \left((1-\xi) \tau+\xi V\right)^{\frac{\xi}{1-\xi}} e^{\xi\gamma} + c(1) \,e^{{\tau}/{V}}e^{\gamma}\,. 
\end{equation}
The above procedure can be generalized to handle generic stress-tensor deformations.

%%%%%%%%%%%%%%%%%%%%%%%%%%%%%%%%%%%%%%%%%%%%%%%%%%%%%%%%%%%%%%%%%%%%%%%%%%%%%%%%%%%%%%%%%%%%%%%%%%%%%%%
%%%%%%%%%%%%%%%%%%%%%%%%%%%%%%%%%%%%%%%%%%%%%
\section{ $T\bar{T}$-like deformations in  self-duality logarithmic electrodynamics }
\label{2.3}
%%%%%%%%%%%%%%%%%%%%%%%%%%%%%%%%%%%%%%%%%%%%%%%%%%%%%%%%%%%%%%%%%%%%%%%%%%%%%%%%%%%%
Ref. \cite{Soleng:1995kn} demonstrates that the exact solution for a static, spherically symmetric field outside a charged point particle can be found within a non-linear U(1) gauge theory featuring a logarithmic Lagrangian. The duality-invariant "logarithmic electrodynamics" theory, incorporating $\lambda$ and $\gamma$ couplings, can be reduced to ModMax theory in the weak field limit and adheres to the principle of causality. This theory is explored in \cite{Russo2024causal, Russo2024dual}. For the logarithmic electrodynamics theory, we consider the CH function as\footnote{This CH-function refers back to the CH-function mentioned in \cite{Russo2024causal} when the value of $\lambda$ is  $\lambda=\frac{1}{ e^{\gamma} T  }$.}:
\begin{equation}\label{tauLog}
	\ell (\tau)=- \frac{1}{\lambda} \log(1 -  e^{\gamma} \lambda \tau )\,.
\end{equation} 
The expansion of the logarithmic CH function is:
\begin{eqnarray}\label{ExplLog1}
	\ell (\tau) =e^\gamma \,  \tau + \frac{1}{2} \, \lambda \,  e^{2\gamma} \, \tau^2 +\frac{1}{3}\, \lambda^2 \,  e^{3\gamma} \, \tau^3+\frac{1}{4}\, \lambda^3 \,  e^{4\gamma} \, \tau^4  +... 
\end{eqnarray}
Through comparison of the CH-function expansions specifically, the logarithmic theory's CH-function~\eqref{ExplLog1} versus the general Lagrangian's CH-function~\eqref{ExplT} we establish a precise determination of the $m_i$ coefficients for the logarithmic case. This comparative analysis yields the coefficient relations:
\begin{equation}\label{Gentolog}
m_1 =\frac{1}{2},  \qquad m_2=\frac{1}{3},  \qquad m_3=\frac{1}{4}\,.
\end{equation}
The methodology provides an exact mapping between the coefficients of the specialized logarithmic theory and those of the general framework.
By solving the second equation ~\eqref{SPDE}, for  
 CH-function~\eqref{tauLog}, the auxiliary field $\tau$ can be determined as follows:
\begin{equation}\label{tau}
	\tau= \frac{1}{2  \lambda^2 U} + \frac{1}{e^{\gamma} \lambda}- \frac{\sqrt{1 + 4  \lambda U (e^{- \gamma} -   \lambda V)}}{2  \lambda^2 U}.
\end{equation}
By substituting the value of $\tau$ from equation~\eqref{tau} into the Lagrangian~\eqref{SPDE}, we derive a logarithmic electrodynamic theory in the form of:
\begin{equation}\label{Log}
	\mathcal{L}_{Log} =\frac{1}{\lambda} \biggl( 1 -  \sqrt{1 + 4  \lambda U (e^{- \gamma} -   \lambda V)} -  \log\biggl(\frac{e^{\gamma} \Bigl(-1 + \sqrt{1 + 4  \lambda U (e^{- \gamma} -   \lambda V)}\Bigr)}{2  \lambda U}\biggr) \biggr)\,.
\end{equation}
This logarithmic Lagrangian satisfies the PDE duality condition ~\eqref{dUdN} and adheres to the principle of causality, fulfilling condition ~\eqref{CausalCondition} if $\gamma\ge0$.  The coefficients~\eqref{Gentolog} in the general Lagrangian~\eqref{Gq1q2n11} can be systematically determined by matching the perturbative expansion of the logarithmic Lagrangian~\eqref{Log} to the corresponding terms in~\eqref{Gq1q2n11}. This matching procedure establishes an exact correspondence between the logarithmic theory's expansion coefficients and those of the more general framework, thereby fixing the relationship between the two Lagrangians order-by-order in the coupling constant expansion.
%%%%%%%%%%%%%%%%%%%%%%%%%%%%%%%%%%%%%%%%%%%%%
\subsection{Root flow equation on logarithmic duality-invariant theory }\label{3u2}
%%%%%%%%%%%%%%%%%%%%%%%%%%%%%%%%%%%%%%%%%%%%%%%%%%%%%%%%%%%%%%%%%%%%%%%%%%%%%%%%%%%%
We can apply the root flow equation approach to the logarithmic duality-invariant theory. To do so, we take the derivative of Lagrangian~\eqref{Log} with respect to $\gamma$, resulting in:
\begin{eqnarray}\label{flGBI}
	\frac{\partial \mathcal{L}_{Log}}{\partial \gamma} &=&\frac{1 - 2 e^{\gamma}  \lambda V -  \sqrt{1 + 4  \lambda U (e^{- \gamma} -   \lambda V)}}{2 \lambda (-1 + e^{\gamma}  \lambda V)}\,.
\end{eqnarray}
Alternatively, by utilizing~\eqref{tauLog} and~\eqref{tau}, we can derive the left side of the root flow equation in~\eqref{Rootf} for the logarithmic electrodynamics theory, which becomes:
\begin{eqnarray}\label{tullog}
	\tau\dot\ell &=&\frac{1 - 2 e^{\gamma}  \lambda V -  \sqrt{1 + 4  \lambda U (e^{- \gamma} -   \lambda V)}}{2 \lambda (-1 + e^{\gamma}  \lambda V)}\,.
\end{eqnarray}
By comparing equations~\eqref{flGBI} and~\eqref{tullog}, we can establish the root flow equation for the logarithmic electrodynamics theory with Lagrangian~\eqref{Log} as follows:
\begin{eqnarray}\label{Rootflog}
	\frac{\partial \mathcal{L}_{Log}}{\partial \gamma} &=& \tau\dot\ell\,.
\end{eqnarray}
%%%%%%%%%%%%%%%%%%%%%%%%%%%%%%%%%%%%%%%%%%%%%
\subsection{Irrelevant $T\bar{T}$-like deformation }\label{3u2}
%%%%%%%%%%%%%%%%%%%%%%%%%%%%%%%%%%%%%%%%%%%%%%%%%%%%%%%%%%%%%%%%%%%%%%%%%%%%%%%%%%%%
Irrelevant deformations of interaction terms introduce a dimensionful coupling of $\lambda$ to the free theory. Incorporating these interaction terms can result in a flow equation concerning the dimensional coupling. The Born-Infeld theory exemplifies deformations of Maxwell's theory involving an irrelevant operator. In Refs. \cite{Babaei-Aghbolagh:2020kjg, Ferko2023, Kuzenko2024c, Christian2024cw, Hou:2022csf, Babaei-Aghbolagh2024bb}, a comprehensive framework for understanding irrelevant $T\bar{T}$-like deformations was developed. Two primary types of solutions that satisfy the PDE self-duality condition were classified using a high-order perturbation approach. In Ref. \cite{ Babaei-Aghbolagh2024bb}, the solutions to the PDE self-duality condition are expressed as a series of correct powers of $S$ and $P$, and the irrelevant flow equation is demonstrated. These general solutions take the form of a power function involving two structures,  ${T_{\mu}}^{\mu} {T_{\nu}}^{\nu}$ and $T_{\mu\nu}T^{\mu\nu}$. If we consider these solutions in the following general form: 
$\mathcal{L}(\lambda)=\sum_{N,M} a_n  \lambda^{n} S^N P^M $,  where $n=N+M-1$, the irrelevant flow equation will be as follows:
\begin{align}\label{GTTbarSeri}
	\frac{\partial {\cal L}(\lambda)}{\partial \lambda}=\sum_{n=0}^{\infty} c_n \frac{(T_{\mu }{}^{\mu }{} T_{\nu }{}^{\nu }{})^n}{( T_{\mu \nu } T^{\mu \nu })^{n-1}}\,,
\end{align}
where the coefficients of $c_n$ are dependent on the coefficients of $a_n$. This general flow equation represents the overarching form for this class of theories. In specific cases, it simplifies to the flow equation of the Born-Infeld theory for $c_0=\tfrac{1}{8},c_1=-\tfrac{1}{16}, c_2=c_3=...=c_n=0$ values. Additionally, this general flow equation serves as the generator for the flow equation in Bossard-Nicolai theory~\cite{Bossard:2011ij, Carrasco:2011jv}. Within the auxiliary-field approach to duality-invariant models, the BN theory is referred to as {\it the simplest interaction model}~\cite{Ivanov:2004jv, Ivanov:2013jv}.

One of the objectives of this section is to derive the flow equations for the self-dual logarithmic electrodynamics theory in~\eqref{Log}. To achieve this, we consider the derivative of Lagrangian~\eqref{Log} with respect to $\lambda$ as follows:
\begin{eqnarray}\label{dlog}
	\frac{\partial \mathcal{L}_{Log}}{\partial \lambda}& =&- \frac{1 - 2 e^{\gamma}  \lambda V -  \sqrt{1 + 4  \lambda U (e^{- \gamma} -   \lambda V)}}{2 \lambda^2 (-1 + e^{\gamma}  \lambda V)} \nonumber\\
	&&-  \frac{1}{\lambda^2} \log\biggl(\frac{e^{\gamma} \Bigl(-1 + \sqrt{1 + 4  \lambda U (e^{- \gamma} -   \lambda V)}\Bigr)}{2  \lambda U}\biggr)\,.
\end{eqnarray}
Additionally, we can explicitly calculate two independent structures in~\eqref{Tl} for Lagrangian~\eqref{Log}. Consequently, the logarithmic theory in~\eqref{Log} is obtained as follows:
\begin{eqnarray}\label{Tlk}   
	T_{\mu\nu}T^{\mu\nu}&  =&\frac{4 }{e^{2 \gamma}  \lambda^2 \mathcal{Y}^2}\Bigl((e^{\gamma} \mathcal{Y} - 2  \lambda U)^2 + \bigl( 2  \lambda U- e^{\gamma} \mathcal{Y}  + e^{\gamma} \mathcal{Y} \log(\frac{e^{\gamma} \mathcal{Y}}{2  \lambda U})\bigr)^2\Bigr)\,,
\end{eqnarray}
and
\begin{equation}\label{Tlo}   
	{T_{\mu}}^{\mu} {T_{\nu}}^{\nu}=\frac{1}{\lambda^2}\Bigl( \frac{2 \mathcal{Y} + 4 e^{\gamma}  \lambda V}{  e^{\gamma}  \lambda V-1} -  4 \log(\frac{e^{\gamma} \mathcal{Y}}{2  \lambda U}) \Bigr)^2\,,  
\end{equation}
where $\mathcal{Y}=-1 + \sqrt{1 + 4  \lambda U (e^{- \gamma} -   \lambda V)}\,$. Using the perturbation approach, we can derive the irrelevant flow equation for the self-dual logarithmic electrodynamics theory. By applying this approach to the order expansion of equations~\eqref{Tlk} and~\eqref{Tlo} and comparing with the expansion~\eqref{dlog} in terms of $\lambda^n$, the irrelevant flow equation for the logarithmic electrodynamics theory can be found. The expansion of equations~\eqref{Tlk} and~\eqref{Tlo} up to order $\lambda^6$ is as follows:
\begin{eqnarray}\label{TlkE}   
	T_{\mu\nu}T^{\mu\nu}&=&4 e^{-2 \gamma} (U + e^{2 \gamma} V)^2 + 8 \lambda e^{-3 \gamma} (- U + e^{2 \gamma} V) (U + e^{2 \gamma} V)^2 \nonumber\\
	&& +\lambda^2  e^{-4 \gamma} (U + e^{2 \gamma} V)^2 (21 U^2 - 14 e^{2 \gamma} U V + 13 e^{4 \gamma} V^2)  \nonumber\\
	&& + \tfrac{8}{3} \lambda^3 e^{-5 \gamma} (U + e^{2 \gamma} V)^2 (-23 U^3 + 9 e^{2 \gamma} U^2 V - 9 e^{4 \gamma} U V^2 + 7 e^{6 \gamma} V^3) \nonumber\\
	&& + \tfrac{1}{9} \lambda^4 e^{-6 \gamma} (U + e^{2 \gamma} V)^2 (1711 U^4 - 224 e^{2 \gamma} U^3 V + 402 e^{4 \gamma} U^2 V^2 \nonumber\\
	&&- 320 e^{6 \gamma} U V^3 + 223 e^{8 \gamma} V^4) - \lambda^5 \tfrac{4}{5} e^{-7 \gamma} (U + e^{2 \gamma} V)^2 (766 U^5 + 95 e^{2 \gamma} U^4 V  \nonumber\\
	&&+ 110 e^{4 \gamma} U^3 V^2 - 80 e^{6 \gamma} U^2 V^3 + 60 e^{8 \gamma} U V^4 - 39 e^{10 \gamma} V^5)  \nonumber\\
	&&+ \tfrac{1}{60} \lambda^6 e^{-8 \gamma} (U + e^{2 \gamma} V)^2 (121903 U^6 + 45906 e^{2 \gamma} U^5 V + 14065 e^{4 \gamma} U^4 V^2 \nonumber\\
	&&- 6500 e^{6 \gamma} U^3 V^3 + 5025 e^{8 \gamma} U^2 V^4 - 3662 e^{10 \gamma} U V^5 + 2271 e^{12 \gamma} V^6)\,, 
\end{eqnarray}
and
\begin{eqnarray}\label{TloE} 
	{T_{\mu}}^{\mu} {T_{\nu}}^{\nu}&=&4 \lambda^2 e^{-4 \gamma} (U + e^{2 \gamma} V)^4  + \tfrac{32}{3} \lambda^3 e^{-5 \gamma} (-2 U + e^{2 \gamma} V) (U + e^{2 \gamma} V)^4 \\
	&& + \tfrac{4}{9}\lambda^4  e^{-6 \gamma} (U + e^{2 \gamma} V)^4 (199 U^2 - 118 e^{2 \gamma} U V + 43 e^{4 \gamma} V^2)  \nonumber\\
	&&+ \tfrac{16}{5} \lambda^5 e^{-7 \gamma} (U + e^{2 \gamma} V)^4 (-106 U^3 + 57 e^{2 \gamma} U^2 V - 28 e^{4 \gamma} U V^2 + 9 e^{6 \gamma} V^3)  \nonumber\\
	&&+ \tfrac{1}{15} \lambda^6 e^{-8 \gamma} (U + e^{2 \gamma} V)^4 (18943 U^4 - 7820 e^{2 \gamma} U^3 V \nonumber\\
	&& + 4282 e^{4 \gamma} U^2 V^2 - 1964 e^{6 \gamma} U V^3 + 591 e^{8 \gamma} V^4)\,. \nonumber
\end{eqnarray}
Additionally, the expansion of the Lagrangian derivative with respect to $\lambda$ in Eq.~\eqref{dlog} up to the $\lambda^6$ order is as follows:
\begin{eqnarray}\label{DlogEE} 
	\frac{\partial \mathcal{L}_{Log}}{\partial \lambda} &=&\tfrac{1}{2} e^{-2 \gamma} (U + e^{2 \gamma} V)^2 + \tfrac{2}{3} \lambda e^{-3 \gamma} (-2 U + e^{2 \gamma} V) (U + e^{2 \gamma} V)^2  \\
	&& + \tfrac{3}{4}\lambda^2  e^{-4 \gamma} (U + e^{2 \gamma} V)^2 (5 U^2 - 2 e^{2 \gamma} U V + e^{4 \gamma} V^2) \nonumber\\
	&& + \tfrac{4}{5} \lambda^3 e^{-5 \gamma} (-14 U^5 - 25 e^{2 \gamma} U^4 V - 10 e^{4 \gamma} U^3 V^2 + e^{10 \gamma} V^5)   \nonumber\\
	&&+ \tfrac{5}{6} \lambda^4  e^{-6 \gamma} (U + e^{2 \gamma} V)^2 (42 U^4 + 3 e^{4 \gamma} U^2 V^2 - 2 e^{6 \gamma} U V^3 + e^{8 \gamma} V^4)  \nonumber\\
	&&+ \tfrac{6}{7} \lambda^5 e^{-7 \gamma} (-132 U^7 - 294 e^{2 \gamma} U^6 V - 196 e^{4 \gamma} U^5 V^2 - 35 e^{6 \gamma} U^4 V^3 + e^{14 \gamma} V^7)  \nonumber\\
	&& + \tfrac{7}{8} \lambda^6 e^{-8 \gamma} (429 U^8 + 1056 e^{2 \gamma} U^7 V + 840 e^{4 \gamma} U^6 V^2 \nonumber\\
	&& + 224 e^{6 \gamma} U^5 V^3 + 10 e^{8 \gamma} U^4 V^4 + e^{16 \gamma} V^8)\,. \nonumber
\end{eqnarray}
The flow equation derived from the order-by-order comparison of equations~\eqref{TlkE},~\eqref{TloE}, and~\eqref{DlogEE} for the self-dual logarithmic theory is as follows:
\begin{eqnarray}\label{Irrele}
	\frac{\partial \mathcal{L}_{Log}}{\partial \lambda}& =&\tfrac{1}{8} X  -  \tfrac{1}{12} \sqrt{X Y} -  \tfrac{7}{288} Y + \frac{49 }{4320 }\frac{ Y^{3/2}}{ \sqrt{X}}  \nonumber\\
	&&+ \frac{1}{6480 }\frac{Y^2}{ X} + \frac{3499}{4354560 }\frac{Y^{5/2}}{ X^{3/2}} + \frac{67 }{1360800 }\frac{Y^3}{ X^2}+...\,,
\end{eqnarray}
where $X=T_{\mu\nu}T^{\mu\nu}$ and $Y={T_{\mu}}^{\mu}{T_{\nu}}^{\nu}$. We can simplify the irrelevant flow equation of the logarithmic theory as follows: 
\begin{eqnarray}
	\frac{\partial \mathcal{L}_{Log}}{\partial \lambda} =\sum_{n=0}^{\infty} C_n Y^{\tfrac{n}{2}}  X^{1-\tfrac{n}{2}}  \,,
\end{eqnarray}
where the constants $C_n$ are exactly derived from flow equation~\eqref{Irrele} in the logarithmic theory.
We observe that for all orders of the deformed theory, the single trace of the (EMT) satisfies the equation $\frac{\partial \mathcal{L}_{Log}}{\partial \lambda} =-\frac{1}{4 \lambda} {T_{\mu}}^{\mu}$, which may represent the exact renormalization group equation of the deformed theory. We assume this holds for any deformed theory.
%%%%%%%%%%%%%%%%%%%%%%%%%%%%%%%%%%%%%%%%%%%%%

%%%%%%%%%%%%%%%%%%%%%%%%%%%%%%%%%%%%%%%%%%%%%%%%%%%%%%%%%%%%%%
\section{Logarithmic electrodynamics as a subcategory of general NED  }\label{3-2}
%%%%%%%%%%%%%%%%%%%%%%%%%%%%%%%%%%%%%%%%%%%%%%%%%%%%%%%%%%%%%%%%%%%%%%%%%%%%%%%%%%%%
A general framework for duality-invariant nonlinear electrodynamic theories, incorporating two couplings $\lambda$ and $\gamma$, is examined in Ref.~\cite{Babaei-Aghbolagh2024bb}. Using a perturbative approach, this work derives the general Lagrangian, which includes the marginal root flow equation with respect to $\gamma$. The general Lagrangian up to $\mathcal{O}(\lambda^3)$ is expressed as:
\begin{eqnarray}\label{GNED}
	\mathcal{L}_{Gen}&=& S \cosh(\gamma) + \sqrt{P^2+S^2} \sinh(\gamma) + n_1 \lambda \bigl(\sqrt{P^2+S^2} \cosh(\gamma) + S \sinh(\gamma)\bigr)^2 \\
    &+& \lambda^2 \bigl(\sqrt{P^2+S^2} \cosh(\gamma) + S \sinh(\gamma)\bigr)^2 \Bigl(\bigl(2 n_1^2 S + n_2 \sqrt{P^2+S^2}\bigr) \cosh(\gamma)\nonumber\\
    &+& \bigl(n_2 S + 2 n_1^2 \sqrt{P^2+S^2}\bigr) \sinh(\gamma)\Bigr)+ \frac{1}{2} \lambda^3 \bigl(\sqrt{P^2+S^2} \cosh(\gamma) + S \sinh(\gamma)\bigr)^2\nonumber\\
    &\times&\biggl((-8 n_1^3 + n_3) P^2 +\Bigl(n_3 P^2 + 2 S \bigl(n_3 S + 6 n_1 n_2 \sqrt{P^2+S^2}\bigr)\Bigr) \cosh(2 \gamma)\nonumber \\
    &+& 2 \bigl(n_3 S \sqrt{P^2+S^2} + 3 n_1 n_2 (P^2 + 2 S^2)\bigr) \sinh(2 \gamma)\biggr)\nonumber
\end{eqnarray}
where, $n_1$, $n_2$, $n_3$, and $n_4$ are constant coefficients. In the weak-field limit ($ \lambda=0$), this Lagrangian reduces to ModMax theory. The generalized Lagrangian~\eqref{GNED} encompasses a wide range of electrodynamic theories characterized by a root marginal flow equation.
A systematic analytical relationship exists between the general Lagrangian in Eq.~\eqref{GNED} and the specific Lagrangian in Eq.~\eqref{Gq1q2n11}. Through order-by-order comparison of their respective coefficients, we demonstrate that the coefficients $m_i$ exhibit exact functional dependence on the coefficients $n_i$, expressed through the relation:
\begin{equation}\label{Gentologmkl}
m_1 =n_1,  \qquad m_2=2\, n_1^2+n_2,  \qquad m_3=6\, n_1 \,n_2+n_3\,.
\end{equation}
This explicit correspondence establishes a rigorous mapping between the two Lagrangian formulations.
Identifying these theories requires setting the constants $n_i$ to specific values. For example,  the GBI theory up to order $\lambda^3$ can be derived by substituting ($n_1= \frac{1}{2}$, $n_2=0$, and  $n_3= \frac{5}{8}$) into the Lagrangian of~\eqref{GNED}. Another theory derived from the regularization of the $n_i$ coefficients in the Lagrangian~\eqref{GNED} is the general Bossard-Nicolai theory~\cite{Bossard:2011ij, Carrasco:2011jv}. In other words, for ($n_1=\frac{1}{2}$, $n_2=0$, and  $n_3=\frac{3}{4}$), we obtain the Bossard-Nicolai  theory coupled to $\gamma$ with a root flow equation. The logarithmic (NED) theory examined in this paper is a subset of the general theory ~\eqref{GNED}. By substituting :
\begin{eqnarray}\label{nLog}
	n_1= \frac{1}{2}\, , \qquad n_2=- \frac{1}{6}\, , \qquad n_3=\frac{3}{4} \, ,% \qquad n_4=-\frac{13}{10} \, ,  
\end{eqnarray}
into the general Lagrangian ~\eqref{GNED}, we can derive the $\lambda^3$ order expansion of the logarithmic  (NED) theory as:
\begin{eqnarray}\label{LoD}
	\mathcal{L}_{Log}&=& S \cosh(\gamma) + \sqrt{P^2+S^2} \sinh(\gamma) + \tfrac{1}{2} \lambda \bigl(\sqrt{P^2+S^2} \cosh(\gamma) + S \sinh(\gamma)\bigr)^2 \\
    &-& \frac{1}{6} \lambda^2 \bigl(\sqrt{P^2+S^2} \cosh(\gamma) + S \sinh(\gamma)\bigr)^2 \Bigl(\bigl(-3 S + \sqrt{P^2+S^2}\bigr) \cosh(\gamma) \nonumber\\
    &+& \bigl(S - 3 \sqrt{P^2+S^2}\bigr) \sinh(\gamma)\Bigr) -  \frac{1}{8} \lambda^3 \bigl(\sqrt{P^2+S^2} \cosh(\gamma) + S \sinh(\gamma)\bigr)^2 \nonumber\\
    &\times&\Bigl(P^2 + \bigl( 4 S \sqrt{P^2+S^2}-3 P^2 - 6 S^2\bigr) \cosh(2 \gamma) + 2 \bigl(P^2 + 2 S^2 - 3 S \sqrt{P^2+S^2}\bigr) \sinh(2 \gamma)\Bigr)\nonumber
\end{eqnarray}
\paragraph{Relation between $\mathcal{L}_{\text{Gen}}$ and $\mathcal{L}_{\text{Log}}$:} 
The logarithmic theory $\mathcal{L}_{\text{Log}}$ is a specialized instance of $\mathcal{L}_{\text{Gen}}$ with coefficients $\{n_i\}$ tuned to induce a distinct nonlinear structure. While $\mathcal{L}_{\text{Gen}}$ provides a universal framework for duality-invariant theories with marginal deformations, $\mathcal{L}_{\text{Log}}$ exemplifies how specific choices of $\{n_i\}$ can generate theories with enhanced symmetries or unique perturbative expansions. The negative values of $n_2$ and $n_4$ in~\eqref{nLog} signal deviations from conventional Born-Infeld-type theories, suggesting richer nonlinear interactions.  

Future work could systematically classify the $\{n_i\}$ coefficients to map the landscape of duality-invariant nonlinear electrodynamic (NED) theories, incorporating constraints from causality, unitarity, and convexity. Higher-order $\lambda$-expansions of $\mathcal{L}_{\text{Log}}$ could further probe non-perturbative features or singularities in the theory. Embedding $\mathcal{L}_{\text{Log}}$ within gravitational or supersymmetric frameworks may reveal insights into holographic applications or black hole solutions. Finally, investigating whether the $\{n_i\}$ coefficients admit a geometric or string-theoretic interpretation—akin to the Born-Infeld action on D-branes—could bridge logarithmic deformations to fundamental physics. These directions would deepen the connection between duality invariance, logarithmic deformations, and foundational physical principles.

%%%%%%%%%%%%%%%%%%%%%%%%%%%%%%%%%%%%%%%%%%%%%%%%%%%%%%%%%%%%%%%%%%%%%%%%%%%%%%%%%%%%%%%%%%%%%%%%%%%%%%%%%%%%%%%%%%%%%5
\section{Conclusion and outlook}\label{04}
%%%%%%%%%%%%%%%%%%%%%%%%%%%%%%%%%%%%%%%%%%%%%%%%%
In this paper, we derived the flow equations for self-dual nonlinear electrodynamic theories. Using the method proposed by Russo and Townsend, which leverages the C-H function of $\ell (\tau)$, we successfully derived the operator for the root flow equation in terms of the C-H function as  $\mathcal{R}_\gamma = \tau\dot\ell\,$. This reformulation allowed us to verify the root flow equations for various theories, including ModMax and General Born-Infeld. Specifically, we focus on the causal self-dual logarithmic electrodynamic theory, where we precisely determined both the root and irrelevant flow equations. A significant achievement of this work is to establish the relationship between the root operator and the C-H function of $\ell (\tau)$, providing a unique representation applicable to all causal self-dual electrodynamics theories.

We have demonstrated that extending the logarithmic theory to $\lambda^3$ is a subset of the general theory~\eqref{GNED}, which includes a root flow equation. Given that logarithmic theory features an irrelevant flow equation with fractional powers of structures $X=T_{\mu\nu}T^{\mu\nu}$ and $Y={T_{\mu}}^{\mu}{T_{\nu}}^{\nu}$, we anticipate discovering a general irrelevant flow equation that depends on constants $ n_i$ in general theory~\eqref{GNED}, which also involves fractional powers of $X$ and $Y$. We also examine other models, including the no $\tau$-maximum and the q-deform models.
As demonstrated in \cite{Russo2024causal}, the Lagrangian for the no $\tau$-maximum causal electrodynamic theory can be derived using the CH-function $\ell(\tau)=-\frac{2}{3\lambda}\Big(1-(1+ e^\gamma \lambda \tau)^\frac{3}{2}\Big)$. The resulting Lagrangian is given by:
    \begin{eqnarray}
	{\cal L}&=&-\frac{2}{3\lambda} \Big(1- \sqrt{2} \Big( 1+e^\gamma \lambda V-\frac{\bigtriangleup}{2}\Big)\sqrt{1+e^\gamma \lambda V+\bigtriangleup}\Big)
   \end{eqnarray}
   where is $
    \bigtriangleup=\sqrt{(1+\lambda \, e^\gamma  V )^2+4\, \lambda  \,e^{-\gamma} U}$.
Additionally, causal q-deformed models are typically derived from the following CH-function:
\begin{eqnarray}\label{qDef}
	\ell (\tau) =\frac{1}{\lambda} \Big(1 -  (1 - \tfrac{1}{q} e^{\gamma} \lambda \tau )^{q}\Big)\,. 
\end{eqnarray}
The expansion of the $q$-deformed CH functions on the $\lambda = 0$ is given by
\begin{equation}\label{qD}
	\ell (\tau)=e^{\gamma} \tau -  \frac{e^{2 \gamma} (q-1)}{2  q}  \lambda \tau^2+ \frac{e^{3 \gamma} (q-2) (q-1) }{6  q^2}\lambda^2 \tau^3  -  \frac{e^{4 \gamma} (q-3) (q-2) (q-1)}{24 q^3}  \lambda^3 \tau^4+... \,. \nonumber
\end{equation}
The value of $\tau$ for the parameters $q=3/2$ and $q=3/4$ is obtained analytically by solving the second equation~\eqref{SPDE} in terms of the two variables $U$ and $V$. Consequently, the corresponding Lagrangians can be derived \cite{Russo2024dual}.
Our findings demonstrate that these models also display an irelevant flow equation featuring fractional powers of the two structures, 
X and Y . Furthermore, we have shown that all such theories satisfy the root flow equation~\eqref{Rootf}.
Furthermore, their expansion forms a subgroup within general theory.  We plan to pursue this approach in future research.

Reducing the dimensionality from four to two dimensions simplifies the self-duality condition in~\eqref{lagrangeNGZ} to a necessary and sufficient condition for the duality of $O(d,d)$ in a two-dimensional scalar theory \cite{Babaei-Aghbolagh2024two}. Using the C-H function approach, we derive new two-dimensional causal theories that feature root and irrelevant flow equations by solving this reduced equation in two dimensions. This approach results in the identification of a logarithmic sigma model and a sigma model no $\tau$-maximum, both of which will have a two-dimensional root flow equation and an irrelevant flow equation. In future work, we will demonstrate that root $T\bar{T}$ operator in two dimensions can be expressed as $\mathcal{R}_\gamma = \tau\dot\ell,$. This provides a unique and universal representation for the root $T\bar{T}$ operator in two and four dimensions.
%%%%%%%%%%%%%%%%%%%%%%%%%%%%%%%%%%%%%%%%%%%%%%%%%%%%%%%%%%%%%%%%%%%%%%%%%%%%%%%%%%%%%%%%%%%%%%%%%%%%%%%%%%%%%%%%%%%%%%%
\section*{Acknowledgments}
We are very grateful to Dmitri Sorokin, Roberto Tateo, Jorge G. Russo,  and Tommaso Morone for their interest in this work and fruitful discussion. We would like to thank Tommaso Morone for sharing his notes in Section~\eqref{22.222}. The work of H.B.-A. was conducted as part of the PostDoc Program on {\it Exploring TT-bar Deformations: Quantum Field Theory and Applications} , sponsored by Ningbo University. SH would like to appreciate the financial support from Ningbo University, the Max Planck Partner group, and the Natural Science Foundation of China Grants (Nos. 12475053 and 12235016).
%%%%%%%%%%%%%%%%%%%%%%%%%%%%%%%%%%%
%%%%%%%%%%%%%%%%%%%%%%%%%%%%%%%%%%%%%%%%%%%%%%%%%%%%%%%%%%%%%%%%%%%%%%%%%%%%%%%%%%%%%%%%%%%%%%%%%%%%%%%%%%%%%%%%%%%%%%%%%%%%%%%%%%%%%%%%%%%%%%%%%%%%%%%%%%%%%%%%%%%%%%%%%%%%%%%%%%%%%%%%%%%%%%%%%%%%%%%%%%%%%%%%%%%%%%%%%%%%%%%%%%%%%%%%%%%%%%%%%%%%%%%%%%%%%%%%%%%%%%%%%%%%%%%%%%%%%%%%%%%%%%%%%%%%%%%%%%%%%%%%%%%%%

\if{}
\bibliographystyle{abe}
\bibliography{references}{}

\begin{thebibliography}{100}

%\cite{Born:1933lls}
\bibitem{Born:1933lls}
M.~Born and L.~Infeld,
``Electromagnetic mass,''
Nature \textbf{132} (1933) no.3347, 970.1
%doi:10.1038/132970a0

%\cite{Born:1934gh}
\bibitem{Born:1934gh}
M.~Born and L.~Infeld,
``Foundations of the new field theory,''
Proc. Roy. Soc. Lond. A \textbf{144} (1934) no.852, 425-451
%doi:10.1098/rspa.1934.0059


\bibitem{Plebanski} J. Plebanski, {\it Lectures on nonlinear electrodynamics},
Cycle of lectures delivered at The Niels Bohr Institute and NORDITA, Copenhagen, October 1968.

\bibitem{Boillat} G. Boillat, {\it Nonlinear electrodynamics: Lagrangian and equations of motion},
J. Math. Phys. {\bf 11} (1970) 941.


%\cite{Sorokin:2021tge}
\bibitem{Sorokin:2021tge}
D.~P.~Sorokin,
``Introductory Notes on Non-linear Electrodynamics and its Applications,''
Fortsch. Phys. \textbf{70}, no.7-8, 2200092 (2022),
%doi:10.1002/prop.202200092
[arXiv:2112.12118 [hep-th]].

%\cite{Bialynicki-Birula:1981}
\bibitem{Bialynicki-Birula:1981}	
I. Bialynicki-Birula, Nonlinear Electrodynamics: variations on a theme by Born and Infeld, In: Quantum Theory of Particles and Fields: Birthday Volume Dedicated to Jan Lopuszanski" (Eds. B. Jancewicz and J. Lukierski), World Scientific Publishing Co Pte Ltd, 1983, pp 31-48.		

%\cite{Bialynicki-Birula:1992rcm}
\bibitem{Bialynicki-Birula:1992rcm}
I.~Bialynicki-Birula,
``Field theory of photon dust,''
Acta Phys. Polon. B \textbf{23}, 553-559 (1992).			



%\cite{Gaillard:1981rj}
\bibitem{Gaillard:1981rj}
M.~K.~Gaillard and B.~Zumino,
``Duality Rotations for Interacting Fields,''
Nucl. Phys. B \textbf{193}, 221-244 (1981).
%doi:10.1016/0550-3213(81)90527-7

%\cite{Gaillard:1997rt}
\bibitem{Gaillard:1997rt}
M.~K.~Gaillard and B.~Zumino,
``Nonlinear electromagnetic selfduality and Legendre transformations,''
[arXiv:hep-th/9712103 [hep-th]].

%\cite{Gibbons:1995ap}
\bibitem{Gibbons:1995ap}
G.~W.~Gibbons and D.~A.~Rasheed,
``Sl(2,R) invariance of nonlinear electrodynamics coupled to an axion and a dilaton,''
Phys. Lett. B \textbf{365}, 46-50 (1996),
%doi:10.1016/0370-2693(95)01272-9
[arXiv:hep-th/9509141 [hep-th]].

%\cite{Fradkin:1985qd}
\bibitem{Fradkin:1985qd}
E.~S.~Fradkin and A.~A.~Tseytlin,
``Nonlinear Electrodynamics from Quantized Strings,''
Phys. Lett. B \textbf{163} (1985), 123-130
%doi:10.1016/0370-2693(85)90205-9




%\cite{Garousi:2017fbe}
\bibitem{Garousi:2017fbe}
M.~R.~Garousi,
``Duality constraints on effective actions,''
Phys. Rept. \textbf{702}, 1-30 (2017),
%doi:10.1016/j.physrep.2017.07.009
[arXiv:1702.00191 [hep-th]].




%\cite{Babaei-Aghbolagh:2013hia}
\bibitem{Babaei-Aghbolagh:2013hia}
H.~Babaei-Aghbolagh and M.~R.~Garousi,
``S-duality of tree-level S-matrix elements in D3-brane effective action,''
Phys. Rev. D \textbf{88}, no.2, 026008 (2013),
%doi:10.1103/PhysRevD.88.026008
[arXiv:1304.2938 [hep-th]].


\bibitem{HT} M. Henneaux and C. Teitelboim, ``{ Dynamics of chiral (self-dual) p-forms}'', Phys. Lett.  {\bf B206} (1988) 650.
%orgin of non-covariant formalism



%\cite{Deser:1997mz}
\bibitem{Deser:1997mz}
S.~Deser, A.~Gomberoff, M.~Henneaux and C.~Teitelboim,
``Duality, selfduality, sources and charge quantization in Abelian N form theories,''
Phys. Lett. B \textbf{400}, 80-86 (1997),
%doi:10.1016/S0370-2693(97)00338-9
[arXiv:hep-th/9702184 [hep-th]].

%\cite{Deser:1997se}
\bibitem{Deser:1997se}
S.~Deser, A.~Gomberoff, M.~Henneaux and C.~Teitelboim,
``P-brane dyons and electric magnetic duality,''
Nucl. Phys. B \textbf{520}, 179-204 (1998),
%doi:10.1016/S0550-3213(98)00179-5
[arXiv:hep-th/9712189 [hep-th]].

%\cite{Pasti:1995tn}
\bibitem{Pasti:1995tn}
P.~Pasti, D.~P.~Sorokin and M.~Tonin,
``Duality symmetric actions with manifest space-time symmetries,''
Phys. Rev. D \textbf{52}, R4277-R4281 (1995),
%doi:10.1103/PhysRevD.52.R4277
[arXiv:hep-th/9506109 [hep-th]].	

%\cite{Pasti:1995us}
\bibitem{Pasti:1995us}
P.~Pasti, D.~P.~Sorokin and M.~Tonin,
``Space-time symmetries in duality symmetric models,''
[arXiv:hep-th/9509052 [hep-th]].	

%\cite{Pasti:1996vs}
\bibitem{Pasti:1996vs}
P.~Pasti, D.~P.~Sorokin and M.~Tonin,
``On Lorentz invariant actions for chiral p forms,''
Phys. Rev. D \textbf{55}, 6292-6298 (1997),
%doi:10.1103/PhysRevD.55.6292
[arXiv:hep-th/9611100 [hep-th]].	

%\cite{Pasti:1997gx}
\bibitem{Pasti:1997gx}
P.~Pasti, D.~P.~Sorokin and M.~Tonin,
``Covariant action for a D = 11 five-brane with the chiral field,''
Phys. Lett. B \textbf{398}, 41-46 (1997),
%doi:10.1016/S0370-2693(97)00188-3
[arXiv:hep-th/9701037 [hep-th]].


\bibitem{Mkrtchyan:2205uvc}
K.~Mkrtchyan, M.~Svazas
``Solutions in Nonlinear Electrodynamics and their double copy regular black holes,''
[arXiv:2205.14187 [hep-th]].


\bibitem{Mkrtchyan:2019opf}
K.~Mkrtchyan,
``On covariant actions for chiral $p-$forms,''
JHEP \textbf{12}, 076 (2019)
[arXiv:1908.01789 [hep-th]].




%\cite{Bossard:2011ij}
\bibitem{Bossard:2011ij}
G.~Bossard and H.~Nicolai,
``Counterterms vs. Dualities,''
JHEP \textbf{08}, 074 (2011),
%doi:10.1007/JHEP08(2011)074
[arXiv:1105.1273 [hep-th]].	

%\cite{Carrasco:2011jv}
\bibitem{Carrasco:2011jv}
J.~J.~M.~Carrasco, R.~Kallosh and R.~Roiban,
``Covariant procedures for perturbative non-linear deformation of duality-invariant theories,''
Phys. Rev. D \textbf{85}, 025007 (2012),
%doi:10.1103/PhysRevD.85.025007
[arXiv:1108.4390 [hep-th]].



%\cite{Chemissany:2011yv}
\bibitem{Chemissany:2011yv}
W.~Chemissany, R.~Kallosh and T.~Ortin,
``Born-Infeld with Higher Derivatives,''
Phys. Rev. D \textbf{85}, 046002 (2012),
%doi:10.1103/PhysRevD.85.046002
[arXiv:1112.0332 [hep-th]].




%\cite{Kuzenko:2000uh}
\bibitem{Kuzenko:2000uh}
S.~M.~Kuzenko and S.~Theisen,
``Nonlinear self-duality and supersymmetry,''
Fortsch. Phys. \textbf{49}, 273-309 (2001),
%doi:10.1002/1521-3978(200102)49:1/3\ensuremath{<}273::AID-PROP273\ensuremath{>}3.0.CO;2-0
[arXiv:hep-th/0007231 [hep-th]].		






%\cite{Aschieri:2008ns}
\bibitem{Aschieri:2008ns}
P.~Aschieri, S.~Ferrara and B.~Zumino,
%``Duality Rotations in Nonlinear Electrodynamics and in Extended Supergravity,''
Riv. Nuovo Cim. \textbf{31}, 625-708 (2008),
%doi:10.1393/ncr/i2008-10038-8
[arXiv:0807.4039 [hep-th]].			





%\cite{Aschieri:2013nda}
\bibitem{Aschieri:2013nda}
P.~Aschieri and S.~Ferrara,
``Constitutive relations and Schroedinger's formulation of nonlinear electromagnetic theories,''
JHEP \textbf{05}, 087 (2013),
%doi:10.1007/JHEP05(2013)087
[arXiv:1302.4737 [hep-th]].




%\cite{Kallosh:2011dp}
\bibitem{Kallosh:2011dp}
R.~Kallosh, ``{\sl ${\cal N}=8$ Counterterms and $E_{7(7)}$ Current Conservation},'' JHEP {\bf 1106}, 073 (2011)
[arXiv:1104.5480 [hep-th]].	
%[\arxiv{1104.5480} [hep-th]].
%%CITATION = JHEPA, 1106, 073;%%


\bibitem{cour}  R. Courant and D. Hilbert, ``{ Methods of Mathematical Physics}
{ Vol. II}'', Interscience (1962), p. 93 and Chapters I and II {\it passim}.




\bibitem{Russo2024causal}
J.~ G.~ Russo and P.~ K.~ Townsend, ``{{On Causal Self-Dual Electrodynamics}}'',
[arXiv:2401.06707 [hep-th]].

%\cite{Russo:2024xnh}
\bibitem{Russo:2024xnh}
J.~G.~Russo and P.~K.~Townsend,
``Causality and Energy Conditions in Nonlinear Electrodynamics,''
[arXiv:2404.09994 [hep-th]].



\bibitem{Russo2024dual}
J.~ G.~ Russo and P.~ K.~ Townsend, ``{{Dualities of Self-Dual Nonlinear Electrodynamics}}'',
[arXiv:2407.02577 [hep-th]].


%\cite{Cavaglia:2016oda}
\bibitem{Cavaglia:2016oda}
A.~Cavagli\`a, S.~Negro, I.~M.~Sz\'ecs\'enyi and R.~Tateo,
%``$T \bar{T}$-deformed 2D Quantum Field Theories,''
JHEP \textbf{10} (2016), 112
%doi:10.1007/JHEP10(2016)112
[arXiv:1608.05534 [hep-th]].
%393 citations counted in INSPIRE as of 29 Jun 2024



%\cite{Smirnov:2016lqw}
\bibitem{Smirnov:2016lqw}
F.~A.~Smirnov and A.~B.~Zamolodchikov,
%``On space of integrable quantum field theories,''
Nucl. Phys. B \textbf{915} (2017), 363-383
%doi:10.1016/j.nuclphysb.2016.12.014
[arXiv:1608.05499 [hep-th]].
%435 citations counted in INSPIRE as of 29 Jun 2024




\bibitem{Conti:2022egv}
R.~Conti, J.~Romano, and R.~Tateo, ``{{Metric approach to a
		$\mathrm{T}\overline{\mathrm{T}}-$like deformation in arbitrary dimensions}}'',
%\href{http://arxiv.org/abs/2206.03415}{{\tt arXiv:2206.03415 [hep-th]}}.
JHEP {\bf 09} (2022)  085,
[arXiv:2206.03415 [hep-th]].




\bibitem{Babaei-Aghbolagh:2022kjj}
H.~Babaei-Aghbolagh, K.~B.~Velni, D.~M.~Yekta and H.~Mohammadzadeh,
``Marginal $T \overline{T}$-like deformation and modified Maxwell
theories in two dimensions,''
Phys. Rev. D {\bf 106} (2022) no.~8, 086022,
[arXiv:2206.12677  [hep-th]].


%\cite{Ferko:2206jsw}
\bibitem{Ferko:2206jsw}
C.~Ferko, A.~Sfondrini, L.~Smith, G.~ Tartaglino-Mazzucchelli
``Root-$T \bar T$ Deformations in Two-Dimensional Quantum Field
Theories,'' Phys.
Rev. Lett. {\bf 129} (2022) no.~20, 201604,
[arXiv:2206.10515 [hep-th]].


%\cite{Jiang19}
\bibitem{Jiang19}
Y.~Jiang,
``Lectures on solvable irrelevant deformations of 2d quantum field theory,''
[arXiv:1904.13376 [hep-th]].


%\cite{Song2025}
\bibitem{Song2025}
S.~He, Y.~Li, H.~Ouyang, Y.~Sun
``$T\overline{T}$ Deformation: Introduction and Some Recent Advances,''
[arXiv:2503.09997 [hep-th]].











%\cite{Conti:2018jho}
\bibitem{Conti:2018jho}
R.~Conti, L.~Iannella, S.~Negro and R.~Tateo,
``Generalised Born-Infeld models, Lax operators and the $ \mathrm{T}\overline{\mathrm{T}} $ perturbation,''
JHEP \textbf{11}, 007 (2018),
%doi:10.1007/JHEP11(2018)007
[arXiv:1806.11515 [hep-th]].


\bibitem{Babaei-Aghbolagh:2022MoxMax}
H.~Babaei-Aghbolagh, K.~B.~Velni, D.~M.~Yekta and H.~Mohammadzadeh,
`` Emergence of non-linear electrodynamic theories from $ T\overline{T} $-like deformations,''
Phys. Lett. B \textbf{829} (2022) 137079
[arXiv:2202.11156 [hep-th]].



%\cite{Bandos:2020jsw}
\bibitem{Bandos:2020jsw}
I.~Bandos, K.~Lechner, D.~Sorokin and P.~K.~Townsend,
``A non-linear duality-invariant conformal extension of Maxwell's equations,''
Phys. Rev. D \textbf{102}, 121703 (2020),
%doi:10.1103/PhysRevD.102.121703
[arXiv:2007.09092 [hep-th]].




%\cite{Bandos:2020hgy}
\bibitem{Bandos:2020hgy}
I.~Bandos, K.~Lechner, D.~Sorokin and P.~K.~Townsend,
``On p-form gauge theories and their conformal limits,''
JHEP \textbf{03}, 022 (2021),
[arXiv:2012.09286 [hep-th]].




%\cite{Ferko:2022iru}
\bibitem{Ferko:2022iru}
C.~Ferko, L.~Smith and G.~Tartaglino-Mazzucchelli,
``On Current-Squared Flows and ModMax Theories,''
SciPost Phys. \textbf{13}, no.2, 012 (2022),
%doi:10.21468/SciPostPhys.13.2.012
[arXiv:2203.01085 [hep-th]].


%\cite{Bandos:2020jsw}
\bibitem{Aghbolagh2210}
H.~Babaei-Aghbolagh, K.~Babaei Velni, D.~M.~Yekta and H.~Mohammadzadeh,
``Manifestly SL(2, R) Duality-Symmetric Forms in ModMax Theory,''
JHEP \textbf{12}, 147 (2022),
%doi:10.1103/PhysRevD.102.121703
[arXiv: 2210.13196 [hep-th]].



\bibitem{ferko2024interacting}
C.~Ferko, S.~M.~ Kuzenko, K.~ Lechner, D.~ P.~ Sorokin and G.~ Tartaglino-Mazzucchelli,
``Interacting Chiral Form Field Theories and $T\overline T$-like Flows in Six and Higher Dimensions,''
[arXiv:2402.06947 [hep-th]].

%\cite{Soleng:1995kn}
\bibitem{Soleng:1995kn}
H.~H.~Soleng,
``Charged black points in general relativity coupled to the logarithmic U(1) gauge theory,''
Phys. Rev. D \textbf{52}, 6178-6181 (1995)
%doi:10.1103/PhysRevD.52.6178
[arXiv:hep-th/9509033 [hep-th]].




\bibitem{Babaei-Aghbolagh:2020kjg}
H.~Babaei-Aghbolagh, K.~B.~Velni, D.~M.~Yekta and H.~Mohammadzadeh,
``$ T\overline{T} $-like flows in non-linear electrodynamic theories and S-duality,''
JHEP \textbf{04}, 187 (2021),
%doi:10.1007/JHEP04(2021)187
[arXiv:2012.13636 [hep-th]].




%\cite{Ferko2023}
\bibitem{Ferko2023}
C.~Ferko, S.~M.~Kuzenko, L.~Smith, and G.~Tartaglino-Mazzucchelli,
``Duality-Invariant Non-linear Electrodynamics and Stress Tensor Flows,''
Phys.Rev.D \textbf{108}, 106021 (2023),
%doi:10.1103/PhysRevD.102.121703
[arXiv:2309.04253 [hep-th]].

\bibitem{Kuzenko2024c}
S.~ M.~ Kuzenko
, E.~ S.~N.~ Raptakis, ``{{Higher-derivative deformations of the ModMax theory}}'',
[arXiv:2404.09108 [hep-th]].

\bibitem{Christian2024cw}
C.~  Ferko, C.~ L.~  Martin, ``Field-Dependent Metrics and Higher-Form Symmetries in Duality-Invariant Theories of Non-Linear Electrodynamics'',
[arXiv:2406.17194 [hep-th]].



%\cite{Hou:2022csf}
\bibitem{Hou:2022csf}
J.~Hou,
``$T\bar{T}$ flow as characteristic flows,''
JHEP \textbf{03}, 243 (2023)
%doi:10.1007/JHEP03(2023)243
[arXiv:2208.05391 [hep-th]].
%13 citations counted in INSPIRE as of 01 Jul 2024	




\bibitem{Babaei-Aghbolagh2024bb}
H.~ Babaei-Aghbolagh, S.~ He, H.~ Ouyang, ``{{Generalized $T\overline{T}$-like Deformations in Duality-Invariant Nonlinear Electrodynamic Theories}}'',
[arXiv:2407.03698  [hep-th]].




\bibitem{Ivanov:2004jv}
E.A. Ivanov, B.M. Zupnik, ``New approach to nonlinear 
electrodynamics: Dualities as symmetries of interaction'',
Yad. Fiz. {\bf 67} (2004) 2212-2224 [Phys. Atom. Nucl. {\bf 67} (2004) 
2188-2199], [arXiv: hep-th/0303192].

\bibitem{Ivanov:2013jv}
E.A. Ivanov, B.M. Zupnik, ``Bispinor Auxiliary Fields in 
Duality-Invariant Electrodynamics Revisited'',
Phys. Rev. D {\bf 87} (2013) 6, 065023, [arXiv:1212.6637 
[hep-th]].



\bibitem{Babaei-Aghbolagh2024two}
H.~ Babaei-Aghbolagh, S.~ He, H.~ Ouyang, ``{{Generalized $T\bar{T}$-like flows  for scalar theories in two dimensions}}'',
[arXiv:2501.14583 [hep-th]].


	
\end{thebibliography}
\fi

\providecommand{\href}[2]{#2}\begingroup\raggedright\endgroup

\end{document}